\documentclass[12pt]{article}
\usepackage[english]{babel}
\usepackage[dvips]{graphicx}
\usepackage{mathtext}

\newcommand{\vth}{\vartheta}
\newcommand{\vph}{\varphi}
\begin{document}

\title{SCALAR FIELD AS DARK ENERGY ACCELERATING EXPANSION OF THE UNIVERSE}
\author{O.~Sergijenko, B.~Novosyadlyj}
\maketitle

\medskip

\centerline{Chair of Theoretical Physics and Astronomical Observatory of} 
\centerline{Ivan Franko National University of Lviv}

\vskip1cm

{\it The features of a homogeneous scalar field $\phi$ with classical Lagrangian $L=\phi_{;i}\phi^{;i}/2-V(\phi)$ and tachyon field Lagrangian $L=-V(\phi)\sqrt{1-\phi_{;i}\phi^{;i}}$ causing the observable accelerated expansion of the Universe are analyzed. The models with constant equation-of-state parameter $w_{de}=p_{de}/\rho_{de}<-1/3$ are studied. For both cases  the fields  $\phi(a)$ and potentials $V(a)$ are reconstucted for the parameters of cosmological model of the Universe derived from the observations. The effect of rolling down of the potential $V(\phi)$ to minimum is shown. } 

\vskip0.5cm

\section*{Introduction}
The cosmological test ``apparent magnitude -- redshift'', realised for  SN Ia \cite{astier2005,astier2006,perlmutter1998,perlmutter1999,riess1998,riess2004,riess2006,wood2007}, and power spectrum of CMB temperature fluctuations, obtained in ground based, stratospheric and cosmic experiments
\cite{Archeops_www,bennett2003,dasi_www,debernardis2000,hanany2000,netterfield2002,hinshaw2007}, surely show that the main part of energy density of the Universe -- more than 70\% --  is dark energy. These and other observable data are satisfactory described by the cosmological model based on the Einstein equations with cosmological constant: $\Omega_{\Lambda}=0.74\pm 0.02$ (see \cite{apunevych2007,novosyadlyj2007,spergel2007} and references therein). But its physical interpretation is rather problematic \cite{carrol2001,peebles1993,peebles1999,sahni2000}. Therefore alternative approaches  --
new physical fields (classical scalar field -- quintessence, tachyon field, k-essence, phantom field, quintom field, Chaplygin gas),
gravity and general relativity modifications, multidimensional gravity, branes and others -- are analyzed (see reviews 
\cite{carrol2001,copeland2006,bludman2006,padmanabhan2003,peebles1999,sahni2000}).

In this paper the scalar field filling the Universe and causing its accelerated expansion is studied. The Universe is considered to be homogeneously filled with the dust matter (density in units of critical $\Omega_m$) and dark energy (density $\Omega_{de}$ and equation of state $-1\le w_{de}\le -1/3$). According to modern data the dust matter consists of cold dark matter ($\Omega_{cdm}\approx 0.21$) and usual baryons ($\Omega_b\approx 0.05$). The reconstruction of scalar field with classical and tachyonic Lagrangian is made for cosmological parameters derived from observations.

\section{Cosmological model and scalar field equations}

We consider the homogeneous and isotropic Universe with metrics of 4-space 
\begin{equation}\label{ds}
ds^2=g_{ij} dx^i dx^j =c^2dt^2-a^2(t)\left[{dr^2\over 1-kr^{2}}+r^2\left(d\vth^2+sin^2\vth d\vph^2\right)\right] ,
\end{equation}
where $k=+1,\;0,\;-1$ for Riemannian, Euclidean and Lobachevskian 3-space respectively. The factor $a(t)$ is a radius of 3-sphere ($k=+1$), 3-pseudosphere ($k=-1$) or the scale factor, normalized  to 1 at current epoch ($k=0$). Here and below the latin indices $i,\,j,\,...$ run from 0 to 3, the greek ones -- over the spatial part of the metrics: $\nu,\, \mu,\,...$=1, 2, 3.
Henceforth we also put $c=1$, so the time variable $t\equiv x_0$ has the dimension of a length. 
If the Universe is filled with dust matter and dark energy, the dynamics of its expansion is completely described by the Einstein equations 
\begin{equation}
R_{ij}-{1\over 2}g_{ij}R=8\pi G \left(T_{ij}^{(m)}+T_{ij}^{(de)}\right),
\label{Einstein-eq}
\end{equation}
where $R_{ij}$ is the Ricci tensor and $T_{ij}^{(m)}$, $T_{ij}^{(de)}$ -- energy-momentum tensors of matter $(m)$ and dark energy $(de)$. If these components interact only gravitationaly then each of them satisfy energy-momentum conservation law separately:
\begin{equation}
T^{i\;\;(m,de)}_{j\;;i}=0
\label{conserv-eq}
\end{equation}
(here and below ``;'' denotes the covariant derivative with respect to the coordinate with given index in space with metrics $g_{ij}$). For ideal fluid with density $\rho_{(m,de)}$ and pressure $p_{(m,de)}$, related by the equation of state $p_{(m,de)}=w_{(m,de)}\rho_{(m,de)}$, it gives
\begin{equation}
\dot{\rho}_{(m,de)}=-3\frac{\dot a}{a} \rho_{(m,de)}(1+w_{(m,de)}) \label{eqconsm}
\end{equation}
(here and below a dot over the variable denotes the derivative with respect to time: ``$\dot{\;\;}$''$\equiv d/dt$). For constant $ w_{(m,de)}$ these equations are easily integrated and 
 dependences of dust matter and dark energy densities on the scale factor are simply obtained:
\begin{equation}
\rho_{m}=\rho_{m}^{(0)}(a/a_0)^{-3},\,\,\,\;\rho_{de}=\rho_{de}^{(0)}(a/a_0)^{-3(1+w_{de})}
\end{equation}
(here and below ``0'' denotes the present values). The matter is considered to be non-relativistic, so  $w_m=0$.

Let the dark energy be a scalar field $\phi({\bf x},t)$ with classical Lagrangian 
\begin{equation}
L=\frac{1}{2}\phi_{;i}\phi^{;i}-V(\phi), \label{lagr_cf}
\end{equation}
where $V(\phi)$ -- potential energy density or the field potential.
We suppose also the scalar field to be homogeneous in expanding homogeneous isotropic Universe  ($\phi({\bf x},t)=\phi(t)$), so its energy density and pressure depends only on time:
\begin{eqnarray}
\rho_{de}(t)=\frac{1}{2}\dot{\phi}^{2}+V(\phi),\,\,\,\,\,
p_{de}(t)=\frac{1}{2}\dot{\phi}^{2}-V(\phi).\label{rho_p}
\end{eqnarray}
Then the conservation law (\ref{conserv-eq}) gives the scalar field evolution equation
\begin{eqnarray}
 \ddot{\phi}+3H\dot\phi=-dV/d\phi, \nonumber
\end{eqnarray}
where $H=\dot{a}/{a}$ is the Hubble constant for any moment of time $t$.

If the Lagrangian of scalar field has the form
\begin{equation}
L=-V(\phi)\sqrt{1-\phi_{;i}\phi^{;i}},\label{lagr_tf}
\end{equation}
such field is called the tachyon one (see \cite{padmanabhan2002} and references therein). The energy density and pressure of the homogeneous tachyon field are defined as follows:
\begin{eqnarray}
\rho_{de}=\frac{V(\phi)}{\sqrt{1-\dot{\phi}^{2}}},\,\,\,\,\,
p_{de}=-V(\phi)\sqrt{1-\dot{\phi}^{2}}.\label{rptach}
\end{eqnarray}
The conservation equation (\ref{conserv-eq}) describes the evolution of the tachyon field in a such way:
\begin{eqnarray*}
\ddot{\phi}+(1-\dot{\phi}^{2})\left(\frac{1}{V}\frac{dV}{d\phi}+3H\dot{\phi}\right)=0.
\end{eqnarray*}
 
Let's define the dimensionless densities $\Omega_m$, $\Omega_{de}$ and $\Omega_{k}$:
$$\Omega_m=\frac{\rho_{m}^{(0)}}{\rho_{cr}^{(0)}},\,\,\,\Omega_{de}=\frac{\rho_{de}^{(0)}}{\rho_{cr}^{(0)}},\,\,\,\Omega_{k}=-\frac{ka_0^{-2}}{H_0^2},$$
where $\rho_{cr}^{(0)}=3H_{0}^{2}/8\pi G$ is critical density of the Universe at the present epoch and $ka_0^{-2}$ -- the 3-space curvature. Using them, the Einstein equations for the model of the Universe with dust matter, dark energy and curvature can be written in form:
\begin{eqnarray}
H&=&H_{0}\sqrt{\Omega_m(a_0/a)^3+\Omega_{k}(a_0/a)^2+\Omega_{de}(a_0/a)^{3(1+w_{de})}},\label{H}\\
q&=&\frac{1}{2}\frac{\Omega_m(a_0/a)^3+(1+3w_{de})\Omega_{de}(a_0/a)^{3(1+w_{de})}}
{\Omega_m(a_0/a)^3+\Omega_{k}(a_0/a)^2+\Omega_{de}(a_0/a)^{3(1+w_{de})}},\label{q}
\end{eqnarray}
where $q=-\ddot{a}/(aH^{2})$ - the acceleration parameter for any time $t$. 

The first equation rewritten for the present time $t_0$ gives the equality: $\Omega_{de}+\Omega_m+\Omega_k=1$. From the second one  it follows that the accelerated expansion ($q<0$) is possible only when $w_{de} < -1/3$.  At the early stages of the evolution of the Universe $q>0$ always and $q\rightarrow 0.5$ when $a\rightarrow 0$. The change of acceleration sign (moment when $q=0$) or transition from deceleration to acceleration occurs at redshift $z_q$ which depends on the relation between the matter and dark energy densities: $z_q=\left[-(1+3w_{de})\Omega_{de}/\Omega_m\right]^{-1/(3w_{de})}-1$.
The scalar field density begins to dominate ($\rho_{de}\ge \rho_m$), when  $z_{de}=\left(\Omega_{de}/\Omega_m\right)^{-1/(3w_{de})}-1$.  To be a candidate for dark energy the scalar field with Lagrangian (\ref{lagr_cf}) must satisfy the condition  $\dot{\phi}^{2}\le V(\phi)$. If  $\dot{\phi}^{2}\ll V(\phi)$ then we will have the scalar field analogous to cosmological constant in the Einstein equations -- $w_{de} = -1$. For field with Lagrangian (\ref{lagr_tf}) the expansion accelerates when $0\le\dot{\phi}^{2}<2/3$.  The scalar field with Lagrangian (\ref{lagr_cf}) which satisfies the conditions $\dot{\phi}^{2}>0$ and $-1\le w_{de}\le -1/3$ is called {\it quintessence} and cosmological models with $\Omega_{de}>\Omega_{cdm}>\Omega_b$ are noted by QCDM. In this work we call QCDM both models (\ref{lagr_cf}) and (\ref{lagr_tf}).
\begin{figure}
\centerline{
\includegraphics[height=6cm]{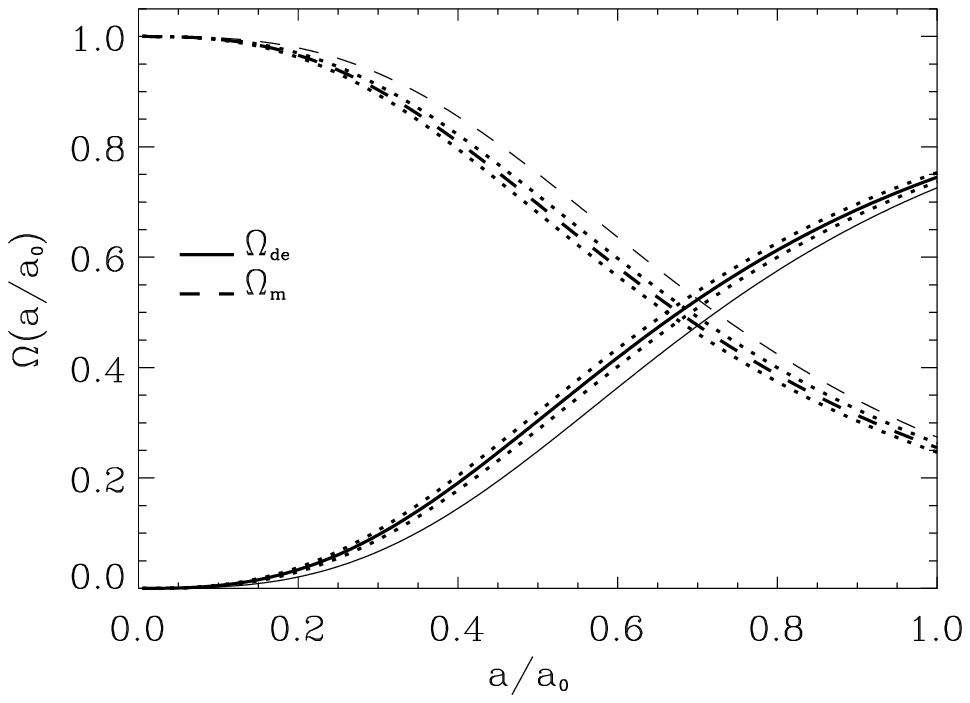}
\includegraphics[height=6cm]{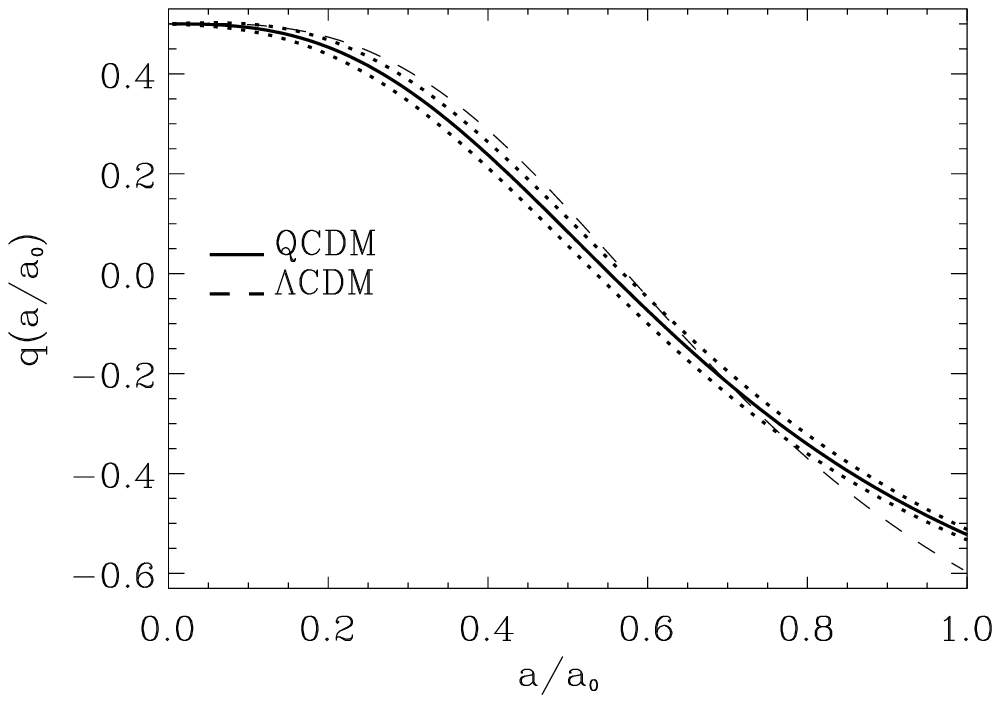}}
\caption{Left: the dark energy (solid line) and matter (dashed) densities (in units of critical density) as functions of the scale factor. The dotted lines show the values calculated at the limits of 1$\sigma$-range of allowed values of the cosmological parameters. For comparison these dependences are also given for the $\Lambda$CDM-model \cite{apunevych2007} (thin lines). Right: the acceleration parameter $q$ as a function of $a$ for the QCDM- and $\Lambda$CDM-models.} 
\label{fig2}
\end{figure}

If $H_{0}$,  $\Omega_m$, $\Omega_{de}$ and $w_{de}$ are derived from the observations of dynamics of the Universe acceleration and CMB temperature fluctuations, it is possible to find the form of $V(\phi)$. The allowed ranges for cosmological parameters determine the uncertanties in the quantities $V$, $\phi$, $q$ and others. We will reconstruct the scalar field for the cosmological model with parameters derived from WMAP and other projects data \cite{spergel2007,wmap_www} (the best-fit values and their 1$\sigma$-confidence intervals):
\begin{eqnarray}
\Omega_{de}=0.745^{+0.017}_{-0.017},\,\,\,
w_{de}=-0.915^{+0.051}_{-0.051},\,\,\,
\Omega_m=0.255^{+0.017}_{-0.017},\,\,\,
h=0.7_{-0.017}^{+0.016}, \label{obs_data}
\end{eqnarray}
where $h\equiv H_0/100$km/s/Mpc. The equation (\ref{H}) for the present time gives $\Omega_{k}=0\pm0.034$.
Fig. \ref{fig2} (left) shows the functions $\Omega_m(a)=\rho_{m}/\rho_{cr}$ and $\Omega_{de}(a)=\rho_{de}/\rho_{cr}$  for the best-fit parameters and their values on upper and lower limits of the confidence intervals (\ref{obs_data}). In this model, as it can be seen in Fig. \ref{fig2}, the matter density is equal to the dark energy one at $z_{de}\approx0.48$ ($a\approx0.68$).  The epoch of the dark energy density domination begins at this redshift. For comparison the corresponding curves for the $\Lambda$CDM-model with parameters \cite{apunevych2007} and $z_{de}\approx0.38$ ($a\approx0.72$) are shown. Their practically identical asymptotic behavior at $a\rightarrow 0$ shows that the dark energy with the parameters (\ref{obs_data}) can't remove the problem of fine tuning of densities of the components  at the end of the phase transitions in the early Universe \cite{linde1990}.
The right panel of Fig. \ref{fig2} shows the acceleration parameter as a function of $a$ for the best-fit  parameters and their values at the upper and lower limits of the confidence intervals (\ref{obs_data}). The acceleration sign change in QCDM-model takes place at $z_{q}\approx0.81$ ($a=0.55$), in $\Lambda$CDM-model -- at $z_{q}\approx0.74$ ($a=0.57$). In QCDM-model (\ref{obs_data}) the main contribution to variation of $q$  at $z\approx 0$ gives the uncertainty of $w_{de}$ and at $z\approx z_{q}$ --  uncertainty of $\Omega_m$.

\section{The evolution of quintessence}

It is possible to find the scalar field $\phi$ and its potential $V$ as the functions of time or the scale factor $a$  using  (\ref{rho_p}) and (\ref{H}):
\begin{eqnarray}
\phi(a)-\phi_{0}&=&\pm\sqrt{\frac{3}{8\pi G}(1+w_{de})}\int_1^{a/a_0}\frac{dy}{y}\times \nonumber\\
&&\sqrt{\frac{\Omega_{de}y^{-3w_{de}}}{1-\Omega_k-\Omega_{de}+\Omega_{k}y+\Omega_{de}y^{-3w_{de}}}},\label{int}\\
V(a)&=&\frac{3H_{0}^{2}}{8\pi G}\Omega_{de}\frac{1-w_{de}}{2}(a_0/a)^{3(1+w_{de})}
\end{eqnarray}
(here and below $\phi_{0}=\phi(a_{0})$). The upper sign corresponds to the field growing in time, the lower sign -- to the decaying one.
Analogous expressions for the growing solution can be found in \cite{elis1991,steinhardt2001,padmanabhan2002,sahni2003}. 
\begin{figure}
\centerline{
\includegraphics[height=6cm]{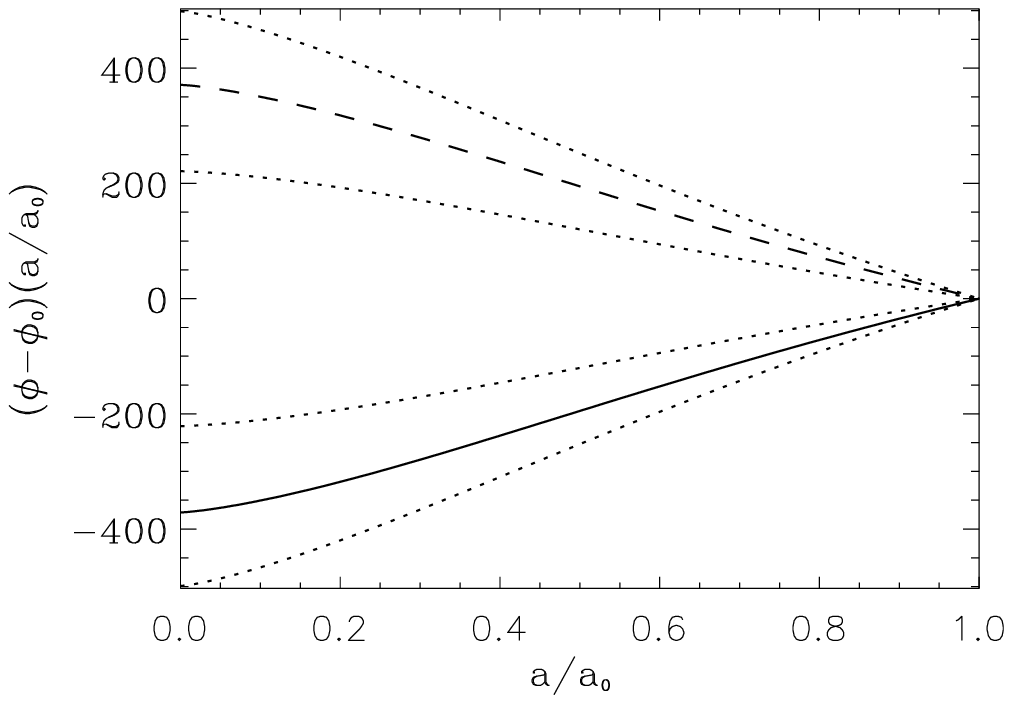}
\includegraphics[height=6cm]{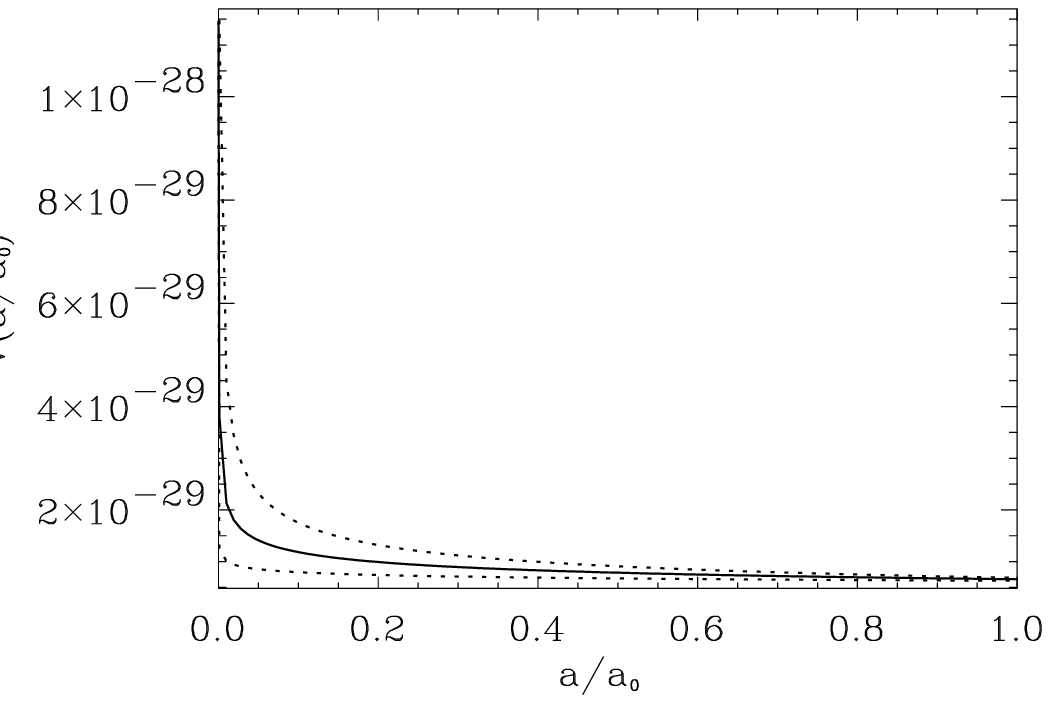}}
\caption{Quintessence $\phi$ (solid line corresponds to the growing, dashed -- to the decaying solution) and its potential $V$ as functions of the scale factor $a$.}
\label{fig4}
\end{figure}
Fig. \ref{fig4} shows the evolution of $\phi(a)$ and $V(a)$ for the best-fit parameters  and their values at the upper and lower limits of the confidence intervals (\ref{obs_data}). It has been found that for small $a$ the most of variation is due to the uncertainty in $w_{de}$ and for $a\sim a_0$ the  uncertainty in $H_{0}$  becomes important. The  uncertainty in $\phi$ is mainly caused by error of $w_{de}$.

When $a\rightarrow0$ then the potential $V$ and kinetic term $\dot{\phi}^2$ go to $\infty$ in a such way  that $w_{de}=(\frac{1}{2}\dot{\phi}^{2}-V)/(\frac{1}{2}\dot{\phi}^{2}+V)=const$. 
For known dependences $\phi(a)$ and $V(a)$ the potential $V(\phi)$ can be presented in a parametric form $(\phi(a),V(a))$. For the flat model ($\Omega_{k}=0$) it is easy to find the explicit function $V(\phi)$. Really, for the model with only two components -- non-relativistic matter and dark energy with $w_{de}=const$ -- integral in (\ref{int}) can be expressed via elementary functions:
\begin{eqnarray}
&&\phi(a)-\phi_{0}=\pm\sqrt{\frac{3}{8\pi G}}\frac{\sqrt{1+w_{de}}}{3w_{de}}\times \nonumber\\ 
&&\ln\left(\frac{\sqrt{(1-\Omega_{de})(a/a_0)^{3w_{de}}+\Omega_{de}}-\sqrt{\Omega_{de}}}{\sqrt{(1-\Omega_{de})(a/a_0)^{3w_{de}}+\Omega_{de}}+\sqrt{\Omega_{de}}}\frac{1+\sqrt{\Omega_{de}}}{1-\sqrt{\Omega_{de}}}\right).\label{phi}
\end{eqnarray}
Inverting it one easily gets:
\begin{eqnarray}
V(\phi-\phi_{0})&=&\frac{3H_{0}^{2}}{8\pi G}\Omega_{de}\frac{1-w_{de}}{2}\left[ch\left(\sqrt{6\pi G}\frac{w_{de}(\phi-\phi_{0})}{\sqrt{1+w_{de}}}\right)\mp\right.  \nonumber \\  &&\left.\frac{1}{\sqrt{\Omega_{de}}}sh\left(\sqrt{6\pi G}\frac{w_{de}(\phi-\phi_{0})}{\sqrt{1+w_{de}}}\right)\right]^{2\frac{1+w_{de}}{w_{de}}}.\label{v}
\end{eqnarray}
So, at the present time such universe is dominated by the classical scalar field with the potential (\ref{v}). The similar expression (for the growing solution) was found also in \cite{sahni2003}.

If $a\rightarrow0$ the field  $\phi(a)$ goes to a finit quantity $\phi_{a=0}$:
\begin{eqnarray*}
\phi_{a=0}-\phi_{0}=\pm\sqrt{\frac{3}{8\pi G}}\frac{\sqrt{1+w_{de}}}{3w_{de}}\ln\left(\frac{1+\sqrt{\Omega_{de}}}{1-\sqrt{\Omega_{de}}}\right),
\end{eqnarray*}
and potential $V$ goes to $\infty$.

In the case of the non-flat 3-space ($\Omega_k\ne 0$) it isn't possible to compute the integral in expression for the scalar field (\ref{int}) analyticaly. But, taking into account the smallness of the quantity $\Omega_{k}$, we represent it as:
\begin{eqnarray}
\phi(a)-\phi_{0}=[\phi(a)-\phi_{0}]_{(k=0)}+\Delta_k^{quin}(a),\label{lin}
\end{eqnarray}
where $[\phi(a)-\phi_{0}]_{(k=0)}$ is (\ref{phi}) and $\Delta_k^{quin}$ is linear in $\Omega_{k}$ correction:
\begin{eqnarray*}
&&\Delta_{k}^{quin}(a)=\mp\sqrt{\frac{3}{8\pi G}}\frac{\sqrt{1+w_{de}}}{2-3w_{de}}\frac{\Omega_{k}\Omega_{de}^{1/2}}{(1-\Omega_{de})^{3/2}}\times\\ &&\left[(a/a_0)^{\frac{2-3w_{de}}{2}}
{_{2}F_{1}}\left(\frac{3}{2},\frac{1}{2}-\frac{1}{3w_{de}};\frac{3}{2}-\frac{1}{3w_{de}};-\frac{\Omega_{de}}{1-\Omega_{de}}(a_0/a)^{3w_{de}}\right)-\right.\\
&&\left.{_{2}F_{1}}\left(\frac{3}{2},\frac{1}{2}-\frac{1}{3w_{de}};\frac{3}{2}-\frac{1}{3w_{de}};-\frac{\Omega_{de}}{1-\Omega_{de}}\right)\right],
\end{eqnarray*}
where ${_{2}F_{1}}(a,b;c;z)\equiv F(a,b;c;z)$ -- hypergeometric function of argument $z$ with parameters $a,\,b,\,c$ \cite{sf}.
It is obvious that now we can't invert $\phi(a)$ and get the explicit expression $V(\phi)$, but the inverse function $\phi(V)$ can be easily written.

Comparison of the results of numerical evaluation of (\ref{int}) with the linear approximation (\ref{lin}) shows, that the error of this approximation is not larger than 0.08\% (at the 1$\sigma$-confidence limits of parameters).  

\begin{figure}
\centerline{
\includegraphics[height=6cm]{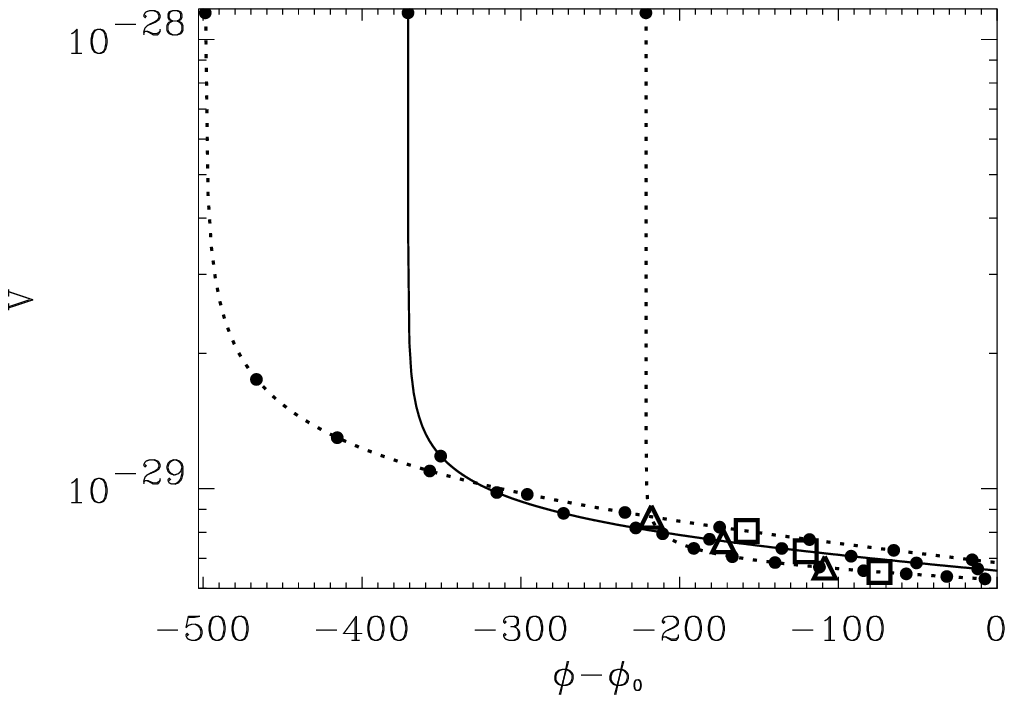}
\includegraphics[height=6cm]{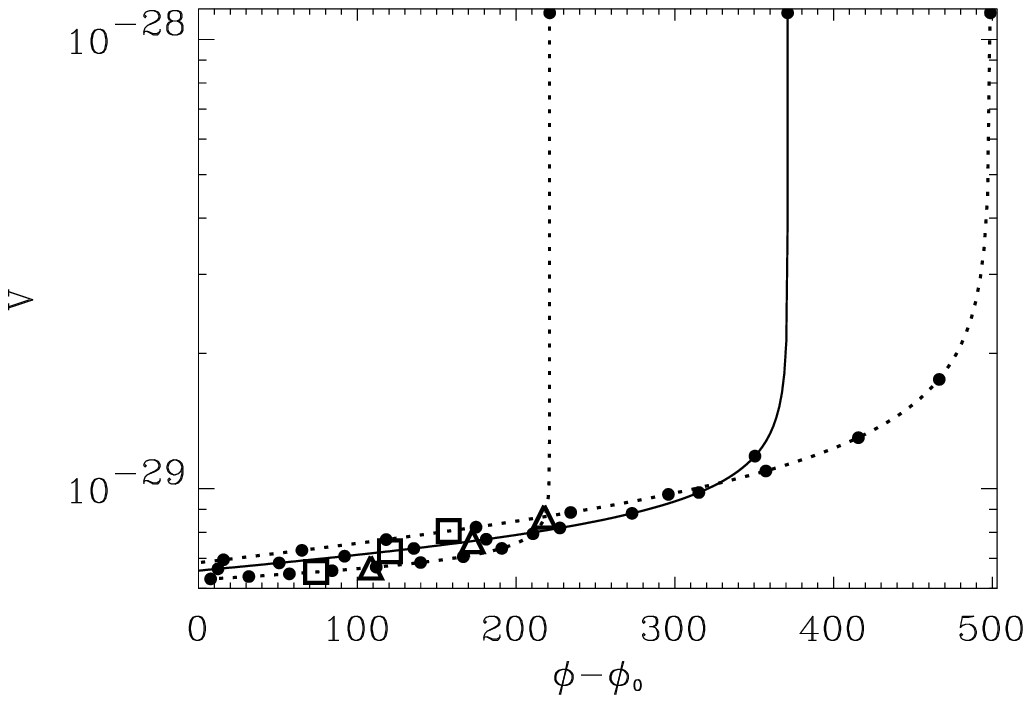}}
\caption{Rolling down of the quintessence which causes the accelerated expansion of the Universe. The left plot corresponds to the field growing in time, the right plot -- to the decaying one. The time is marked along the curves by points in the interval $0.1a_0$ of the scale factor: the left end points correspond to $t\approx0$, the right end points -- the moment which is close to present (left), the right end points correspond to $t\approx0$, the left end points -- to the moment which is close to present (right). Triangles denotes the moments when the sign of the acceleration changes  ($z_q$), squares -- the moments of the matter and dark energy densities equality ($z_{de}$).}
\label{fig6}
\end{figure}
The potentials $V(\phi)$ for the best-fit parameters $w_{de}$, $\Omega_m$, $\Omega_{de}$, $H_{0}$ and their values at the upper and lower limits of the confidence intervals are shown in Fig. \ref{fig6} (time intervals are marked along the curves by points). It can be seen that 2 independent symmetric with respect to $\phi-\phi_0=0$ potentials exist. The shape of the curves suggests that here we have the phase transition -- rolling  down of the field $\phi$ to the minimum $V(\phi)=0$ which is located at $\phi\rightarrow\pm\infty$  ($a\rightarrow\infty$). It has been found that for $\phi(a\approx 0)$ the difference is maximal for the curves with the different values of $w_{de}$, when for $\phi(a\sim a_0)$ the variation of $V(\phi)$ is caused mainly by the uncertainty in  $H_{0}$.

\section{Evolution of the tachyon field}

For the scalar field with Lagrangian (\ref{lagr_tf}) and $w_{de}=\dot\phi^2-1=const$ it can be seen that
$\phi-\phi_0=\sqrt{1+w_{de}}(t-t_0)$: the tachyon field is proportional to the time. However, it is convenient to represent the field $\phi$ and its potential $V$ as functions of the scale factor $a$ or redshift $z$. For given $\Omega_{de}$, $\Omega_m$, $w_{de}$ and $H_{0}$ using (\ref{rptach}) and (\ref{H}) one can find $\phi(a)$ and $V(a)$:
\begin{eqnarray}
\phi(a)-\phi_{0}&=&\pm\frac{\sqrt{1+w_{de}}}{H_{0}}\times\nonumber\\
&&\int_1^{a/a_0}\frac{dy\sqrt{y}}{\sqrt{1-\Omega_k-\Omega_{de}+\Omega_{k}y+\Omega_{de}
y^{-3w_{de}}}}\label{k-t},\\
V(a)&=&\frac{3H_{0}^{2}}{8\pi G}\sqrt{-w_{de}}\Omega_{de}(a_{0}/a)^{3(1+w_{de})}.\label{inttach}
\end{eqnarray}
(similar expressions for the growing solution can be found in \cite{padmanabhan2002,bagla2003}). 
\begin{figure}
\centerline{
\includegraphics[height=6cm]{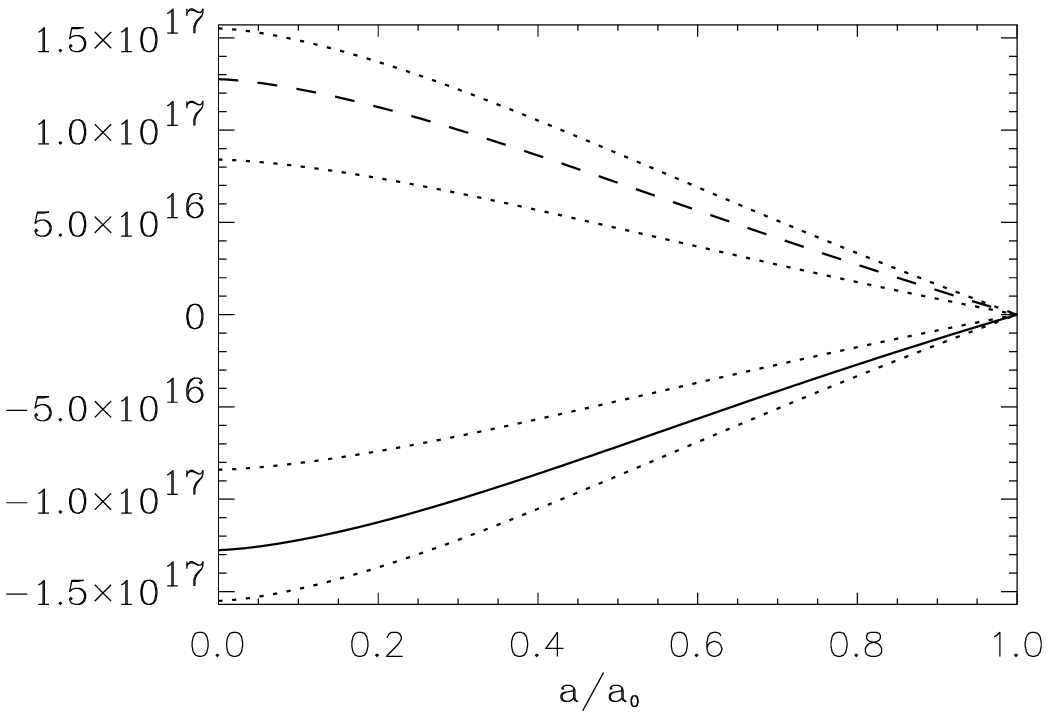}
\includegraphics[height=6cm]{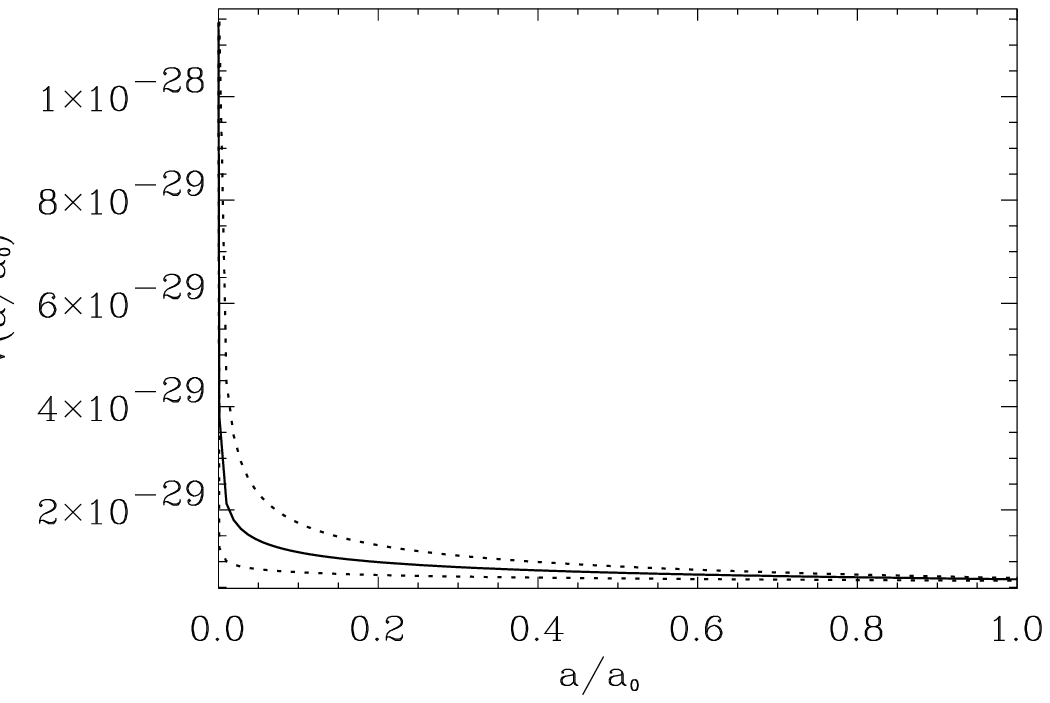}}
\caption{The tachyon field $\phi$ (solid line corresponds to the growing, dashed -- to the decaying solution) and its potential $V$ as functions of the scale factor $a$. }
\label{fig9}
\end{figure}
The dependences $\phi$ and $V$ on $a$ for the best-fit cosmological parameters and their values at the upper and lower limits of $1\sigma$-confidence intervals are  shown in Fig. \ref{fig9}. It has been found that for small $a$ the variation of the potential is mainly caused by the uncertainty  in  $w_{de}$  and for $a\sim a_0$ the related to the  uncertainty in $H_{0}$ one becomes significant. The uncertainty in $w_{de}$ is also important for the variation of $\phi$.

For $\Omega_{k}=0$ we can get $\phi(V)$:
\begin{eqnarray}
&&\phi(V)-\phi_{0}=\pm\frac{2}{3H_{0}}\frac{\sqrt{1+w_{de}}}{\sqrt{1-\Omega_{de}}}\left[\left(\frac{3H_{0}^{2}}{8\pi G}\frac{\sqrt{-w_{de}}\Omega_{de}}{V}\right)^{\frac{1}{2(1+w_{de})}}\times\right.\nonumber \\
&&\left.{_{2}F_{1}}\left(\frac{1}{2},-\frac{1}{2w_{de}};1-\frac{1}{2w_{de}};-\left(\frac{3H_{0}^{2}}{8\pi G}\right)^{-\frac{w_{de}}{1+w_{de}}}\frac{\Omega_{de}^{\frac{1}{1+w_{de}}}}{1-\Omega_{de}}\frac{V^{\frac{w_{de}}{1+w_{de}}}}{(-w_{de})^{\frac{w_{de}}{2(1+w_{de})}}}\right)-\right. \nonumber \\
&&\left.{_{2}F_{1}}\left(\frac{1}{2},-\frac{1}{2w_{de}};1-\frac{1}{2w_{de}};-\frac{\Omega_{de}}{1-\Omega_{de}}\right)\right]. \label{phi-V}
\end{eqnarray}
When $a\rightarrow0$ the field $\phi$ goes to a finit value $\phi_{a=0}$:
\begin{eqnarray*}
\phi_{a=0}-\phi_{0}=\mp\frac{2}{3H_{0}}\sqrt{\frac{1+w_{de}}{1-\Omega_{de}}}{_{2}F_{1}}\left(\frac{1}{2},-\frac{1}{2w_{de}};1-\frac{1}{2w_{de}};-\frac{\Omega_{de}}{1-\Omega_{de}}\right).
\end{eqnarray*}

If $\Omega_{k}\neq0$, the integral (\ref{k-t}) can't be solved analyticaly, but analogically to quintessence it can be written as follows:
\begin{eqnarray}
\phi(V)-\phi_{0}=[\phi(V)-\phi_{0}]_{(k=0)}+\Delta_k^{tach}(V),\label{pkt}
\end{eqnarray} 
where $[\phi(V)-\phi_{0}]_{(k=0)}$ is (\ref{phi-V}) for the flat model and linear in $\Omega_{k}$ correction is
\begin{eqnarray*}
&&\Delta_{k}^{tach}(V)=\mp\frac{1}{5H_{0}}\frac{\Omega_{k}}{1-\Omega_{de}}\sqrt{\frac{1+w_{de}}{1-\Omega_{de}}}\left[\left(\frac{3H_{0}^{2}}{8\pi G}\frac{\sqrt{-w_{de}}\Omega_{de}}{V}\right)^{\frac{5}{6(1+w_{de})}}\times\right.\\
&&\left.{_{2}F_{1}}\left(\frac{3}{2},-\frac{5}{6w_{de}};1-\frac{5}{6w_{de}};-\left(\frac{3H_{0}^{2}}{8\pi G}\right)^{-\frac{w_{de}}{1+w_{de}}}\frac{\Omega_{de}^{\frac{1}{1+w_{de}}}}{1-\Omega_{de}}\frac{V^{\frac{w_{de}}{1+w_{de}}}}{(-w_{de})^{\frac{w_{de}}{2(1+w_{de})}}}\right)-\right.\\
&&\left.{_{2}F_{1}}\left(\frac{3}{2},-\frac{5}{6w_{de}};1-\frac{5}{6w_{de}};
-\frac{\Omega_{de}}{1-\Omega_{de}}\right)\right].
\end{eqnarray*}
Comparing the results of the numerical calculation of (\ref{k-t}) with (\ref{pkt}), we see that the error of the linear in curvature approximation is not larger than 0.08\% for tachyon field too (at the 1$\sigma$-confidence limits of parameters).

\begin{figure}
\centerline{
\includegraphics[height=6cm]{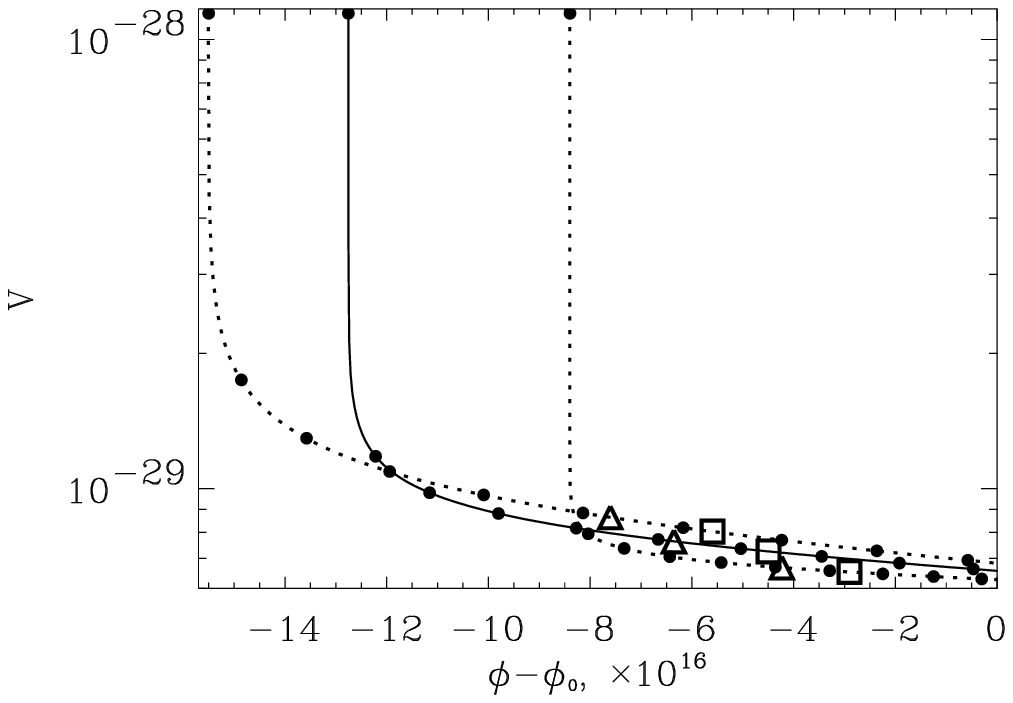}
\includegraphics[height=6cm]{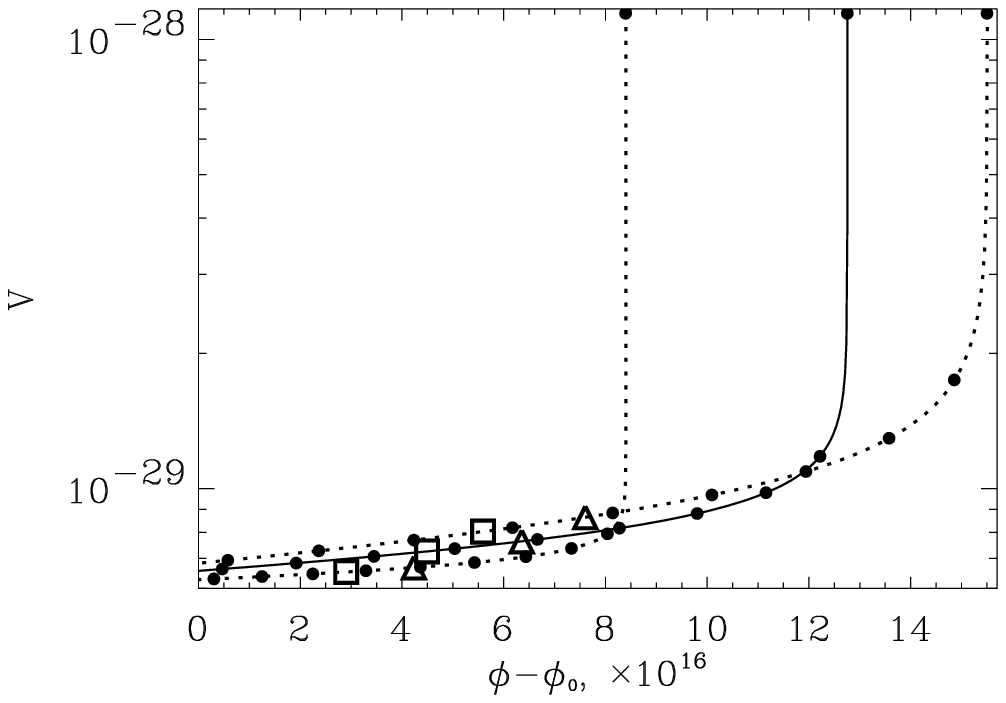}}
\caption{Rolling down of the tachyon field which causes the accelerated expansion of the Universe. The left plot corresponds to the field growing in time, the right plot -- to the decaying one. The time is marked along the curves by points in the interval $0.1a_0$ of the scale factor: the left end points correspond to $t\approx0$, the right end points -- the moment which is close to present (left), the right end points correspond to $t\approx0$, the left end points -- to the moment which is close to present (right). Triangles denotes the moments when the sign of the acceleration changes  ($z_q$), squares -- the moments of the matter and dark energy densities equality ($z_{de}$).}
\label{fig10}
\end{figure}
The potentials $V(\phi-\phi_{0})$ for the best-fit parameters $w_{de}$, $\Omega_m$, $\Omega_{de}$, $H_{0}$ (\ref{obs_data}) and their values at the upper and lower limits are shown in Fig. \ref{fig10}. In this case 2 independent solutions symmetrical with respect to $\phi-\phi_0=0$ also exist. The shape of the curves suggests that here we have the phase transition -- rolling down of the field $\phi$ to the minimum $V(\phi)=0$ which corresponds to $\phi\rightarrow\pm\infty$ at $a\rightarrow\infty$ too. It has been found that uncertainties of the reconstrution of the tachyon field are caused by uncertainties in $w_{de}$, $\Omega_{de}$, $H_0$ similarly to the classical scalar field.

\section{The particle and event horizons}

Let's discuss the size of the observable part of the Universe. 
The causally conected region which can be in principle seen at a given time is delimited by a particle horizon. From the equation of the light cone $ds^{2}=0$, on which light propagates, we find that the rate at which the radius $r$ of a wavefront in metrics (\ref{ds}) for $\vph=\vth=const$ changes is
\begin{eqnarray*}
\frac{dr}{dt}=\frac{\sqrt{1-kr^{2}}}{a(t)}
\end{eqnarray*}
and the physical distance traveled by light from the Big Bang ($t=0$) to the time $t$  (the particle horizon radius) is
\begin{eqnarray*}
R_{p}(t)=a(t)\int^{r(t)}_{0}\frac{dr}{\sqrt{1-kr^{2}}}=a(t)\int^{t}_{0}\frac{dt'}{a(t')}=a\int^{a}_{0}\frac{da'}{a'^{2}H(a')}.
\end{eqnarray*}
The event horizon delimits the region from which we can ever (up to $t=\infty$) receive information about events taking place at time $t$:
\begin{eqnarray*}
R_{e}(t)=a(t)\int^{\infty}_{t}\frac{dt'}{a(t')}=a\int^{\infty}_{0}\frac{da'}{a'^{2}H(a')}-R_{p}(a).
\end{eqnarray*}
Taking into account the equation (\ref{H})  these integrals explicitly read
\begin{eqnarray}
R_{p}(a)=\frac{a}{a_0H_{0}}\int^{a/a_0}_{0}\frac{dy}{\sqrt{y}\sqrt{1-\Omega_k-\Omega_{de}+\Omega_k y+\Omega_{de}y^{-3w_{de}}}} \label{h-p}
\end{eqnarray}
for the particle horizon and 
\begin{eqnarray}
R_{e}(a)=\frac{a}{a_0H_{0}}\int^{\infty}_{0}\frac{dy}{\sqrt{y}\sqrt{1-\Omega_k-\Omega_{de}+\Omega_k y+\Omega_{de}y^{-3w_{de}}}}-R_{p}(a) \label{h-e}
\end{eqnarray}
for the event horizon.

In the case of the flat 3-space it is possible to obtaine the analytical expressions for the particle horizon radius
\begin{eqnarray*}
R_{p}=\frac{2(a/a_0)^{3/2}}{H_{0}\sqrt{1-\Omega_{de}}}{_{2}F_{1}}\left(\frac{1}{2},-\frac{1}{6w_{de}};1-\frac{1}{6w_{de}};-\frac{\Omega_{de}}{1-\Omega_{de}}(a_0/a)^{3w_{de}}\right)
\end{eqnarray*}
and for the event horizon radius
\begin{eqnarray*}
R_{e}=&-&\frac{a/a_0}{3H_{0}\sqrt{(1-\Omega_{de})\pi}w_{de}}\left(\frac{\Omega_{de}}{1-\Omega_{de}}\right)^{\frac{1}{6w_{de}}}\Gamma\left(\frac{1}{2}+\frac{1}{6w_{de}}\right)\Gamma\left(-\frac{1}{6w_{de}}\right)\\
&-&\frac{2(a/a_0)^{3/2}}{H_{0}\sqrt{1-\Omega_{de}}}{_{2}F_{1}}\left(\frac{1}{2},-\frac{1}{6w_{de}};1-\frac{1}{6w_{de}};-\frac{\Omega_{de}}{1-\Omega_{de}}(a_0/a)^{3w_{de}}\right).
\end{eqnarray*}
For $\Omega_k\neq0$ the integrals (\ref{h-p}) and (\ref{h-e}) can't be solved analyticaly, so analogically to the fields we present them in the following form:
\begin{eqnarray}
R_p&=&R_{p(k=0)}+\Delta_k^{p}, \label{h1}\\
R_e&=&R_{e(k=0)}+\Delta_k^{e}, \label{h2}
\end{eqnarray}
where $R_{p(k=0)}$ and $R_{e(k=0)}$ correspond to case of $\Omega_k=0$. The linear in curvature corrections are
\begin{eqnarray*}
\Delta_k^{p}&=&-\frac{(a/a_0)^{5/2}\Omega_k}{3H_0(1-\Omega_{de})^{3/2}}{_{2}F_{1}}\left(\frac{3}{2},-\frac{1}{2w_{de}};1-\frac{1}{2w_{de}};-\frac{\Omega_{de}}{1-\Omega_{de}}(a_0/a)^{3w_{de}}\right),\\
\Delta_k^{e}&=&\frac{(a/a_0)\Omega_k}{3H_0w_{de}\sqrt{\pi}(1-\Omega_{de})^{3/2}}\left(\frac{\Omega_{de}}{1-\Omega_{de}}\right)^{\frac{1}{2w_{de}}}\Gamma\left(-\frac{1}{2w_{de}}\right)\Gamma\left(\frac{3}{2}+\frac{1}{2w_{de}}\right)+\\
&&\frac{(a/a_0)^{5/2}\Omega_k}{3H_0(1-\Omega_{de})^{3/2}}{_{2}F_{1}}\left(\frac{3}{2},-\frac{1}{2w_{de}};1-\frac{1}{2w_{de}};-\frac{\Omega_{de}}{1-\Omega_{de}}(a_0/a)^{3w_{de}}\right).
\end{eqnarray*}
Comparison of the results of the numerical computation of (\ref{h-p})-(\ref{h-e}) with (\ref{h1})-(\ref{h2}) shows that in the linear in $\Omega_k$ approximation the error is not larger than 0.04\% for the particle horizon and 0.05\% for the event horizon (at the 1$\sigma$-confidence limits of parameters).

\begin{figure}
\centerline{\includegraphics[height=6cm]{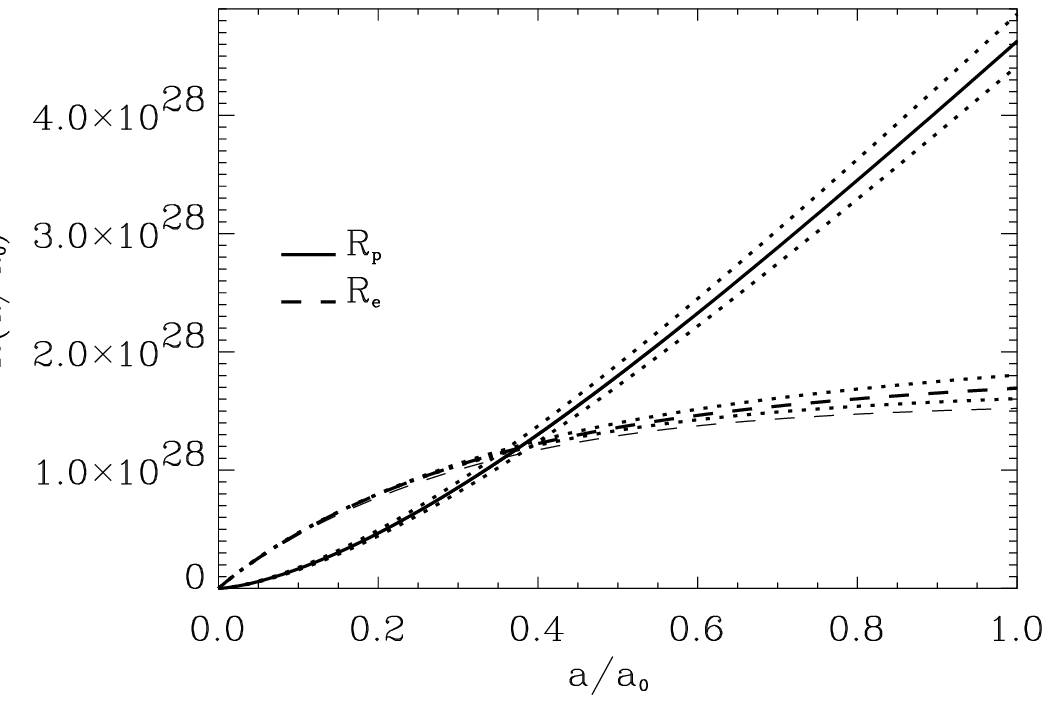}
\includegraphics[height=6cm]{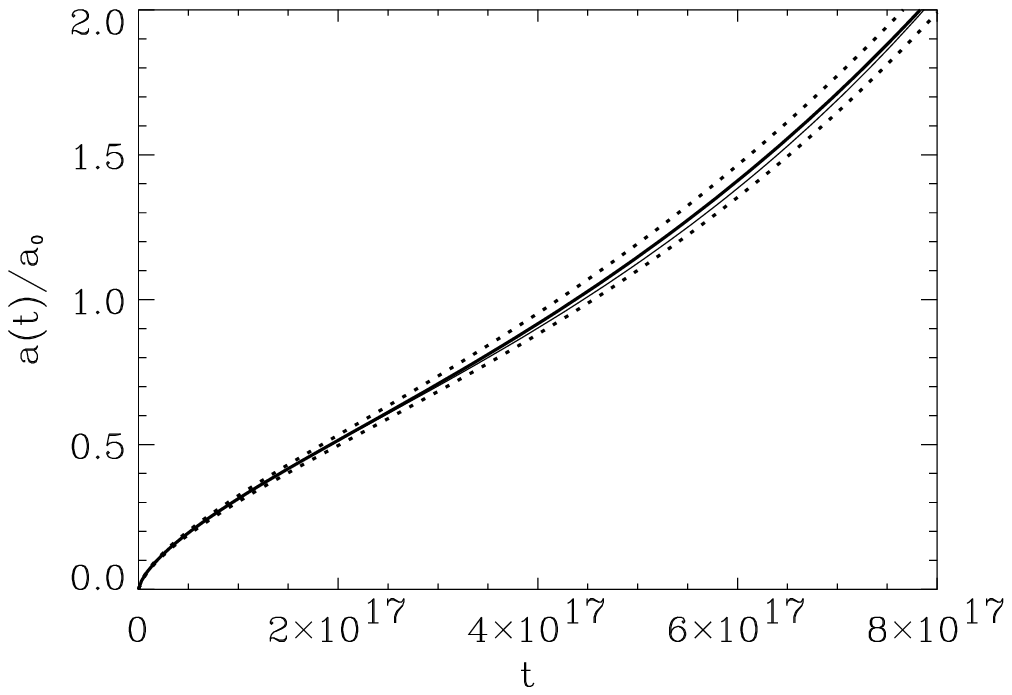}}
\caption{Left: The radii of the particle (solid line) and event (dashed) horizons as functions of the scale factor. Right: The temporal evolution of the scale factor. For comparison the corresponding dependences for the $\Lambda$CDM-model are shown by thin lines.}
\label{fig7}
\end{figure}
In the left panel of Fig. \ref{fig7} $R_{p}(a)$ and $R_{e}(a)$ are plotted for the best-fit cosmological parameters and their values at the upper and lower limits of $1\sigma$-confidence intervals. For the particle horizon the main uncertainty is caused by uncertainties in the Hubble constant and matter density,  for the event horizon -- by the uncertainty in $w_{de}$. The plots for the $\Lambda$CDM-model (thin lines) are also shown for comparison (the lines $R_{p}(a)$ for both models practically overlap).

The fact that at the present epoch the event horizon is smaller than the particle horizon means that the size of the observable today part of the Universe which is determined by the quantity $R_{p}(a)$ refers really to earlier time moments when the given region was situated inside the event horizon (the part of Fig. \ref{fig7} for which $a\le 0.35$ or $z\ge 1.86$).

Let's study the temporal evolution of the scale factor. It is obvious that
\begin{eqnarray}
t=\int^{a}_{0}\frac{da'}{a'H(a')}=\frac{1}{H_{0}}\int^{a/a_0}_{0}\frac{dy\sqrt{y}}{\sqrt{1-\Omega_k-\Omega_{de}+\Omega_k y+\Omega_{de}y^{-3w_{de}}}}.\label{ta}
\end{eqnarray}
The explicit function $t(a)$ in the case of zero 3-space curvature is
\begin{eqnarray}
t=\frac{2(a/a_0)^{3/2}}{3\sqrt{1-\Omega_{de}}H_{0}}{_{2}F_{1}}\left(\frac{1}{2},-\frac{1}{2w_{de}};1-\frac{1}{2w_{de}};-\frac{\Omega_{de}}{1-\Omega_{de}}(a_0/a)^{3w_{de}}\right).\label{t-k-0}
\end{eqnarray}
For $\Omega_k \ne 0$ the integral (\ref{ta}) can't be solved analyticaly, but when $\Omega_k \ll 1$ it is conveniently to present
\begin{equation}
t=t_{(k=0)}+\Delta_k^t,\label{dt}
\end{equation} 
where $t_{(k=0)}$ is the exact expression (\ref{t-k-0}) for the flat 3-space model. Linear in $\Omega_k$ correction is
\begin{eqnarray*}
\Delta_k^t=-\frac{(a/a_0)^{5/2}\Omega_k}{5H_0(1-\Omega_{de})^{3/2}}{_{2}F_{1}}\left(\frac{3}{2},-\frac{5}{6w_{de}};1-\frac{5}{6w_{de}};-\frac{\Omega_{de}}{1-\Omega_{de}}(a_0/a)^{3w_{de}}\right).
\end{eqnarray*}
Comparing the linear approximation with the results of the numerical integration of (\ref{ta}), it is easy to see that in this case the error is not greater than 0.07\% (at the 1$\sigma$-confidence limits of parameters).

The right panel of Fig. \ref{fig7} shows the scale factor as a function of time for the best-fit cosmological parameters and their values at the upper and lower limits of  $1\sigma$-confidence intervals. The main variation in $a(t)$ for $t\ge 2\cdot10^{17}$s is caused by the uncertainty in $H_{0}$. It can be easily seen that for small $t$ the scale factor grows as $a\sim t^{2/3}$ (decelerated expansion) and later (at $t\approx 2\cdot10^{17}$s) the (quasi)exponential accelerated expansion begins -- $a\sim \exp{Ht}$.

As it can be seen from Fig. \ref{fig7}, in the early Universe ($a\le 0.35$, $z\ge 1.86$) the event horizon radius is larger than the particle horizon one both in the $\Lambda$CDM- and QCDM-models. But in the later epoch ($a\ge 0.35$, $z\le 1.86$) $R_p(a)\ge R_e(a)$. At the present epoch ($t_0\approx 14$ billion years) $R_p^{(0)}\approx 15$ Gpc, $R_e\approx 5.5$ Gpc. For comparison, in the QCDM-model with parameters (\ref{obs_data}) the distance to  quasars with $z=2$ is $\approx 5.3$ Gpc and to the CMB last scattering surface is $\approx$14.5 Gpc. It means that the events which take place on the constant-time $t=t_0$ hypersurface at distance $r>R_e$ will never be seen by a terrestrial observer. So, the quasars with $z\ge 2$ have emitted the detected today light when the age of the Universe  was $t\le 3.5$ billion years. When in future the terrestrial astronomers will study the objects on the constant-time $t=t_0$ hypersurface, 
the observable today quasars with $z> 2$ will be already outside the particle horizon and will never ''reenter'' it.

\section{CMB temperature and matter density fluctuations power spectra}

\begin{figure}
\centerline{
\includegraphics[height=6cm]{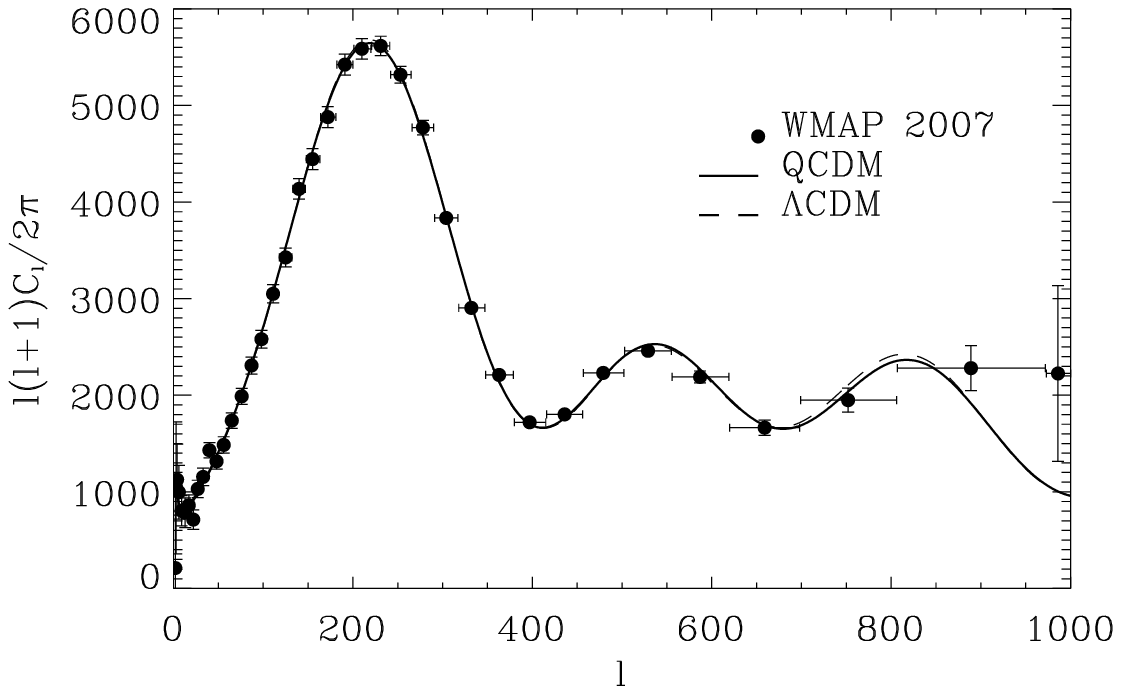}
\includegraphics[height=6cm]{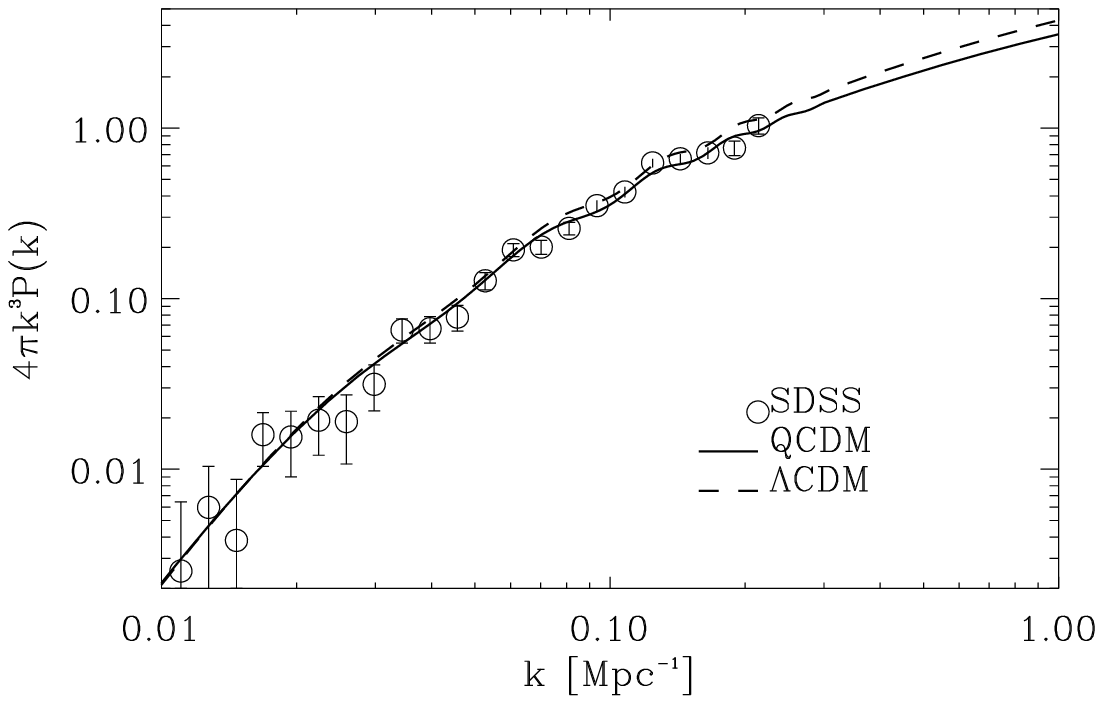}}
\caption{The CMB temperature (left) and matter density (right) fluctuations power spectra for the models of Universe with dark energy: quintessence (solid line), cosmological constant (dashed). }
\label{cl_dk}
\end{figure}

We have analyzed the features of quintessence and tachyon field for the cosmological parameters derived from the observations of the CMB temperature fluctuations and inhomogeneities of the galaxies spatial distribution power spectra and the dynamics of expansion of the Universe. Fig. \ref{cl_dk} shows the observed (circles) and computed power spectra for model with dark energy in the form of scalar field (\ref{lagr_cf}), (\ref{lagr_tf}) and cold dark matter (QCDM) and, for comparison, with cosmological term and cold dark matter ($\Lambda$CDM). The parameters of the QCDM-model are taken from \cite{spergel2007,wmap_www} ($\Omega_{de}=0.745\pm0.017$,
$w_{de}=-(0.915\pm 0.051)$, $\Omega_m=0.255\pm0.017$, $\Omega_{b}h^2=0.02198\pm0.017$, $h=0.7\pm0.017$, $\sigma_8=0.711\pm0.04$, $n_s=0.942\pm0.016$) and of the $\Lambda$CDM-model -- from \cite{apunevych2007} ($\Omega_{\Lambda}=0.736\pm0.065$, $\Omega_m=0.278\pm0.080$, $\Omega_{b}=0.05\pm0.011$,  $h=0.68\pm0.09$, $\sigma_8=0.73\pm0.08$, $n_s=0.96\pm0.015$). As we can see, the curves of CMB temperature fluctuations power spectra for the QCDM- and $\Lambda$CDM-models almost overlap and well approximate the observable data from the WMAP experiment \cite{hinshaw2007}. The difference is maximal near the third acoustic peak  ($\ell\sim 700-800$). The goodness of the approximation of the observable spectrum by a model is characterised by the quantity $\chi^2$:  for the QCDM-model $\chi^2=35.0$, for the $\Lambda$CDM-model $\chi^2=37.2$. Such difference isn't significant for 39 observable points, but suggests that the QCDM-model with the given parameters is a bit closer to a model of the observable Universe. The higher precision of data in the region near the third acoustic peak which is expected in the planned experiment PLANCK would allow us to make a choice between the $\Lambda$CDM- and QCDM-model. In the case of matter density perturbations power spectra the situation is similar to that of CMB one. The unfilled circles in the right panel of the figure show the galaxies concentration perturbations power spectrum from the digital sky survey SDSS \cite{tegmark2004a}. The matter density perturbations power spectra for the $\Lambda$CDM-  and  QCDM-models are simply related to it as $P_{SDSS}(k)=b^2P(k)$, where $b$ is the bising parameter determined by minimization of $\chi^2$. For the $\Lambda$CDM-model $b=1.2$ with $\chi^2_{min}=25.1$, for the QCDM-model $b=1.3$ with $\chi^2_{min}=24.0$.  The difference of $\chi^2_{min}$ in these two models is not significant for 22 observable points. 

Thus, the coincidence of the QCDM-model with observations is a little better than of the $\Lambda$CDM-model, but the difference of $\chi^2_{min}$-criterion isn't sufficient. 

\section*{Conclusion} 

In Fig. \ref{fig2}, \ref{fig7} it is shown that the dynamics of expansion of the Universe and its horizons are almost the same for the $\Lambda$CDM- and QCDM-models with the parameters derived from the same observable data. The differences of the corresponding quantities in the $\Lambda$CDM- and QCDM-models  are comparable to the uncertainties caused by the experimental data errors. At redshifts $z>5$ the dynamics of expansion of the Universe becomes the same as for the standard CDM-model: $\Omega_m=1$, $q=1/2$. It means that for identification of the source causing the accelerated expansion of the Universe (cosmological constant or dark energy) the errors of data of observable cosmology must be reduced approximately by an order. It is shown also that the cosmological effects of the quintessence and tachyon field can be indistinguishable. The accelerated expansion of the Universe for these dark energy models is caused by slow rolling down of the fields $\phi$ to their minima $V\rightarrow0$ at $\phi\rightarrow\pm\infty$ (Fig. \ref{fig6}, \ref{fig10}). For both fields $\phi-\phi_0$ is finit when $a\rightarrow 0$, but  $V(a\rightarrow0)\rightarrow\infty$. For the QCDM-model of the Universe with parameters (\ref{obs_data}) the evolution of both fields is calculated (fig. \ref{fig4}, \ref{fig9}).  The quintessence has in this model such features: $V(a)=3695(a_0/a)^{0.255}$eV/cm$^3$, $\dot\phi^2/V=0.0888$, the tachyon field -- $V(a)=3691(a_0/a)^{0.255}$eV/cm$^3$, $\dot\phi^2=0.085$. The age of the Universe in the QCDM-model is  $t_0\approx 14$ billion years, the present particle horizon -- $R_p\approx 15$ Gpc, the event horizon --  $R_e\approx 5.5$ Gpc. 

\vskip1cm

\section*{Acknowledgements}

This work was supported by the project of Ministry of Education and Science of Ukraine ``The linear and non-linear stages of evolution of the cosmological perturbations in models of the multicomponent Universe with dark energy'' (state registration number 0107U002062) and the research program of National Academy of Sciences of Ukraine ``The exploration of the structure and components of the Universe, hidden mass and dark energy (Cosmomicrophysics)'' (state registration number 0107U007279). The authors are thankful also to Yu. Kulinich for useful discussions.

\end{document}